\documentclass[12pt,twocolumn]{iopart}

\usepackage{graphicx,epsfig,cite}
\usepackage{bm,subfigure,iopams}

\begin{document}

\title[]{Fluctuation theorems and atypical trajectories}

\author{M Sahoo\footnote[1]{Present address: Max Planck Institute of Colloids and Interfaces, Department of Theory and Bio-Systems, Research Campus, Potsdam-Golm, D-14424 Potsdam.}, S Lahiri and A M Jayannavar}

\address{
Institute of Physics, Sachivalaya Marg, Bhubaneswar, Orissa, India - 751005 \eads{\mailto{sahoo@mpikg.mpg.de}, \mailto{lahiri@iopb.res.in}, \mailto{jayan@iopb.res.in}}
}

\newcommand{\nwc}{\newcommand}
\nwc{\para}{\paragraph}
\nwc{\vs}{\vspace}
\nwc{\hs}{\hspace}
\nwc{\la}{\langle}
\nwc{\ra}{\rangle}
\nwc{\del}{\partial}
\nwc{\lw}{\linewidth}
\nwc{\nn}{\nonumber}

\nwc{\pd}[2]{\frac{\partial #1}{\partial #2}}
\nwc{\zprl}[2]{{\it Phys. Rev. Lett.} ~{\bf #1} ~#2}
\nwc{\zpre}[2]{{\it Phys. Rev. E} ~{\bf #1} ~#2}
\nwc{\zpra}[2]{{\it Phys. Rev. A} ~{\bf #1} ~#2}
\nwc{\zjsm}[1]{{\it J. Stat. Mech.} ~#1}
\nwc{\zepjb}[2]{{\it Eur. Phys. J. B} ~{\bf #1} ~#2}
\nwc{\zrmp}[3]{Rev. Mod. Phys. ~{\bf #1},~#2~(#3)}
\nwc{\zepl}[3]{Europhys. Lett. ~{\bf #1},~#2~(#3)}
\nwc{\zjsp}[2]{{\it J. Stat. Phys.}~{\bf #1}~#2}
\nwc{\zptps}[3]{Prog. Theor. Phys. Suppl. ~{\bf #1},~#2~(#3)}
\nwc{\zpt}[2]{{\it Physics Today} ~{\bf #1}~#2}
\nwc{\zap}[2]{{\it Adv. Phys.} ~{\bf #1}~#2}
\nwc{\zjpcm}[2]{{\it J. Phys. Condens. Matter}~{\bf #1}~#2}
\nwc{\zjpa}[2]{{\it J. Phys. A}~{\bf #1}~#2}
\nwc{\zpjp}[3]{Pramana J. Phys. ~{\bf #1},~#2~(#3)}

\begin{abstract}
In this work, we have studied simple models that can be solved analytically to illustrate various fluctuation theorems. These fluctuation theorems provide symmetries individually to the distributions of physical quantities like the classical work ($W_c$), thermodynamic work ($W$), total entropy ($\Delta s_{tot}$) and dissipated heat ($Q$), when the system is driven arbitrarily out of equilibrium. All these quantities can be defined for individual trajectories. We have studied the number of trajectories which exhibit behaviour unexpected at the macroscopic level. As the time of observation increases, the fraction of such atypical trajectories decreases, as expected at macroscale. Nature of distributions for the thermodynamic work and the entropy production in nonlinear models may exhibit peak (most probable value) in the atypical regime without violating the expected average behaviour. However, dissipated heat and classical work exhibit peak in the regime of typical behaviour only. 
\end{abstract}



\maketitle



\section{Introduction}
The last two decades have observed a crescendo of research activity in the field of nonequilibrium statistical mechanics \cite{bus05}. One of the major breakthroughs has been the emergence of the so-called fluctuation theorems \cite{bus05,eva02,har07,rit03,rit06,kur07,eva93,eva94,gal95,jar97,zon02,zon04,nar04,cro99,sei05,sei08,kur98,leb99,jar00,boc77,boc81,boc81a}. These are one of the few relations that are valid even when one is far away from equilibrium, a region that is beyond the scope of the well-established linear response theory. They provide a quantitative measure of the probability of a phase space trajectory to dissipate heat, work or entropy as compared to the probability of its time-reversed trajectory to absorb the same. In other words, it quantifies the probability of violating the average trend which is dictated by the second law. The general relation can be written in the form
\[
\frac{P_f(x)}{P_r(-x)}=e^{\alpha x},
\]
where $\alpha$ is a constant with inverse dimension of that of $x$, $P_f(x)$ and $P_r(x)$ are the probability densities of $x$ along the forward and the backward processes, respectively. As a specific example, let us consider the Crooks' Fluctuation Theorem (CFT) for the dissipated work $W_d$ \cite{cro99,cro98}. $W_d$ is related to the thermodynamic work $W$ and free energy difference $\Delta F$ between terminal states as $W_d=W-\Delta F$. Suppose that a Brownian particle is initially in thermal equilibrium with a heat bath and it is driven into a nonequilibrium state by an external protocol $\lambda(t)$. The dissipated work depends on the microstate of the initial equilibrium state and on the stochastic path of the trajectory as well as on the final $\lambda(t)$. Hence repeated measurements yield different values for $W_d$, the probability of which obeys
\begin{equation}
\frac{P_f(W_d)}{P_r(-W_d)}=e^{\beta W_d},
\end{equation}
$\beta$ being the inverse temperature of the bath. 

The theorem says that the frequency of observing a forward trajectory holds a ratio of $e^{\beta W_d}$ with that of observing a time-reversed one. In macroscopic regime, since dissipated work increases with system size, the above statement implies that there is only small probability for an observer to detect a reverse trajectory. This is consistent with the emergence of macroscopic irreversibility (an outcome of the second law). 

The CFT was an example of what are collectively called the transient fluctuation theorems, for which the system must begin in a state of thermal equilibrium with the bath and is thereafter  allowed to evolve under the given protocol. Using the CFT, the Jarzynski Equality follows \cite{jar97,cro98}, namely,
\begin{equation}
\la e^{-\beta W}\ra = e^{-\beta\Delta F},
\label{JE}
\end{equation}
where $\la\cdots\ra$ denotes ensemble average over all trajectories for a system being prepared in an initial equilibrium state and to the same protocol. It may be noted that the LHS of (\ref{JE}) contains nonequilibrium properties whereas the RHS contains equilibrium free energy difference between the two terminal states. This identity has gained importance due to its ability to calculate the free energy difference from nonequilibrium measurements. If instead of thermodynamic work, we consider the classical work defined below, then the Bochkov-Kuzovlev identity follows \cite{boc77,boc81,boc81a}, i.e.,
\begin{equation}
\la e^{-\beta W_c}\ra = 1.
\label{BKR}
\end{equation}

If the total potential including external protocol is given by $U(x,t)$ (inclusive approach), then the thermodynamic work equals $W=\int_0^\tau \pd{U(x,t)}{t}dt$. If the potential is decomposed as an unperturbed potential $U_0(x)$ and perturbing potential $U_p(x)$ (exclusive approach), then the classical work $W_c$ defined over a time interval $\tau$ is given by $W_c=-\int_0^\tau \pd{U_p(x,t)}{x}\dot x(t)dt$. Seifert has generalized \cite{sei05,sei08} the concept of entropy to a stochastic trajectory. He has proved the integral fluctuation theorem, namely,
\begin{equation}
\la e^{-\Delta s_{tot}}\ra = 1.
\label{IFT}
\end{equation}
Here the average is over the ensemble of finite time trajectories. For this theorem to hold, the initial state of the system need not be in equilibrium. There are also steady state fluctuation theorems where the system begins and thereafter remains throughout in some given (nonequilibrium) steady state. One such theorem is that provided by Seifert \cite{sei05,sei08} for the total entropy:
\begin{equation}
\frac{P(\Delta s_{tot})}{P(-\Delta s_{tot})}=e^{\Delta s_{tot}},
\label{SSFT}
\end{equation}
where $\Delta s_{tot}$ is the total change in entropy of the system and the bath. Similar steady state fluctuation theorems can also be derived for work and heat.
Once again, the connection with the second law is obvious. It can be shown from the above theorems, using Jensen's inequality, that \emph{on average}, we will retrieve the statements of the second law: $\la W\ra\ge\Delta F$ or $\la\Delta s_{tot}\ra\ge 0$. Similarly, we also obtain $\la W_c\ra\ge 0$, as expected for macroscopic systems. By separate analysis, one can show that the average heat $\la Q\ra$ dissipated into the bath follows $\la Q\ra\ge 0$. Here too, given a stochastic trajectory, of a system, one can evaluate the dissipated heat $Q$ using the framework of stochastic thermodynamics \cite{sek}. There are several trajectories which do not obey the properties dictated by the second law, i.e., some trajectories exhibit excursions away from the typical behaviour, $W_c<0$, $W<\Delta F$, $Q<0$ and $\Delta s_{tot}<0$ \cite{eva02a}. It turns out that these trajectories are necessary for the satisfaction of the fluctuation theorems.

Now one might ask: what is the rate of decay of the number of these atypical trajectories as we make the size of the system larger or observe it over a longer period of time \cite{mam10}? We shall show here analytically that if the probability distribution of the observable is Gaussian, then the number decays according as the complementary error function. Further, if the various observables , namely the thermodynamic work ($W$), classical work ($W_c$) and the dissipated heat ($Q$), are measured for each trajectory, then the realizations along which $W<\Delta F$ need not be the same as those along which $W_c<0$, $Q<0$ or $\Delta s_{tot}< 0$. In other words, the atypical trajectories corresponding to one observable may be typical with respect to the others. We have also analyzed the nature of distributions $P(W)$, $P(W_c)$, $P(Q)$ and $P(\Delta s_{tot})$. Surprisingly, the distributions for thermodynamic work and the entropy production in nonlinear models may exhibit most probable value in the atypical regime without violating the expected average behaviour. However, the dissipated heat and the classical work exhibit peak in the regime of typical behaviour only. For simplicity, throughout our analysis, the system is driven out of equilibrium by the same sinusoidal force only.

\section{Harmonic potential}

\subsection{Thermodynamic work distributions}

We consider the following system: a Brownian particle is placed in a harmonic potential and is in contact with a thermal bath at temperature $T$. The system is first allowed to equilibrate with the bath, and thereafter subjected to a given time dependent protocol $f(t)$. The bath is assumed to be infinite and hence always stays at equilibrium. The system, however, goes out of equilibrium. We will consider a sinusoidal perturbation: $f(t)=A\sin(\omega t+\phi)$. The particle follows the overdamped Langevin equation:
\begin{equation}
\gamma \dot x = -kx+f(t)+\xi(t),
\label{oLang}
\end{equation}
where the noise is white and Gaussian, so that $\la\xi(t)\ra=0$ and $\la\xi(t)\xi(t')\ra=2\gamma T\delta(t-t')$. $k$ is the spring constant.

%
%
%
The solution of (\ref{oLang}) is
\begin{eqnarray}
x(t)=x_0\exp(-kt)+\frac{e^{-kt/\gamma}}{\gamma}\int_0^t dt' e^{kt'/\gamma}[A\sin(\omega t'+\phi)+\xi(t')].\nn\\
\end{eqnarray}
Here, $x_0$ is the initial position of the particle, which is supposed to be sampled from a thermal distribution: $P(x_0)=\sqrt{\frac{k\beta}{2\pi}}\exp[-\frac{k\beta}{2}(x_0-\frac{A}{k}\sin\phi)^2]$. The above equation shows that the position variable is linear in $\xi(t)$ and hence is destined to follow a Gaussian distribution as well. The thermodynamic work is defined as \cite{jar97}
\begin{equation}
W(\tau)=\int_0^\tau dt ~x(t)\frac{df(t)}{dt},
\label{W}
\end{equation}
which says that even the work follows a Gaussian distribution.
Thus we can write
\begin{equation}
P(W)=\frac{1}{\sqrt{2\pi\sigma^2}}\exp\left[-\frac{(W-\la W\ra)^2}{2\sigma^2}\right]
\label{Wdist}
\end{equation}
It can be readily shown that the variance is given by
\begin{equation}
\sigma^2\equiv\la W^2\ra-\la W\ra^2=2T(\la W\ra-\Delta F),
\label{FDT}
\end{equation}
where $\Delta F$ is the change in equilibrium free energy, given by
\begin{equation}
\Delta F=-\frac{A^2}{2k}\sin^2(\omega\tau+\phi)+\frac{A^2}{2k}\sin^2\phi.
\end{equation}
Here, $\tau$ is the time of observation. The relation (\ref{FDT}) is referred to as the fluctuation-dissipation theorem. 

Using the relation (\ref{W}) and substituting this in (\ref{Wdist}), one readily obtains the analytical expression for $P(W)$ in terms of the system parameters. The expression for $\la W\ra$ has been given in \ref{appA}.

\subsection{Classical work distributions}

The classical work done by the system is defined through \cite{boc77,jar06,jar07}
\begin{equation}
W_c=\int_0^\tau f(t)\dot x dt = W-f(\tau)x(\tau)+f(0)x(0).
\label{Wc}
\end{equation}
Once again, it can be trivially shown to be Gaussian and is given by
\begin{equation}
P(W_c) = \frac{1}{2\pi\sigma^2}\exp\left[-\frac{(W_c-\la W_c\ra)^2}{2\sigma_c^2}\right]
\end{equation}
with $\sigma_c^2=2T\la W_c\ra$. The value of $\la W_c\ra$ has been given in \ref{appA}. 

\subsection{Entropy distributions}

Entropy is generally considered as an ensemble property. But Seifert has generalized the concept of entropy to a single stochastic trajectory. The total entropy production ($\Delta s_{tot}$) along a single trajectory involves both the system entropy ($\Delta s$) and the entropy change in the medium ($\Delta s_m$). The change in medium entropy is given by
\begin{equation}
\Delta s_m = \frac{Q}{T},
\end{equation}
where $Q$ is the heat dissipated into the bath, which can be readily calculated for a stochastic trajectory generated by the Langevin equation. The nonequilibrium entropy of the system is defined as \cite{sei05,sei08}.
\begin{equation}
s(t) \equiv -\ln P(x(t),t),
\end{equation}
where $P(x,t)$ is the position probability density of the particle. It immediately follows that the change in system entropy for any trajectory of duration $\tau$ is given by
\begin{equation}
\Delta s = -\ln\frac{P(x,\tau)}{P(x_0)},
\end{equation}
where $P(x,\tau)$ and $P(x_0)$ are the probability densities of the particle positions at time $t=0$ and at final time $\tau$, respectively. Thus for a given trajectory $x(t)$, the system entropy $s(t)$ depends on the initial and final probability densities, hence containing information about the whole ensemble. The total entropy change over a time interval $\tau$ is given by
\begin{equation}
\Delta s_{tot} = \Delta s_m + \Delta s.
\end{equation}
$\Delta s_{tot}$ obeys the integral fluctuation theorem \cite{sei05,sei08}:
\begin{equation}
\la e^{-\Delta s_{tot}}\ra =1.
\end{equation}
For our problem, the total entropy becomes 
\begin{equation}
\Delta s_{tot} = \frac{W-\Delta U}{T}-\ln\frac{P(x,\tau)}{P(x_0)}.
\label{stot}
\end{equation}
The first term on the RHS is obtained by replacing $Q$ by $W-\Delta U$ which follows from the first law. Here $\Delta U$ is the change in the internal energy of the system, which is given by
\begin{eqnarray}
\Delta U = U(x,\tau)-U(x_0) 
= \frac{1}{2}kx^2-xf(\tau)-\frac{1}{2}kx_0^2+x_0f(0),
\end{eqnarray}
where $f(\tau)=A\sin(\omega\tau+\phi)$ and $f(0)=A\sin\phi$. Initially the system is prepared in thermal equilibrium given by
\begin{equation}
P(x_0) = \sqrt{\frac{k}{2\pi T}}~\exp\left[-\frac{k(x_0-\la x_0\ra)^2}{2T}\right].
\end{equation}
The evolution of the system through Langevin dynamics leads to the final distribution
\begin{equation}
P(x,\tau) = \sqrt{\frac{k}{2\pi T}}~\exp\left[-\frac{k(x-\la x\ra)^2}{2T}\right].
\end{equation}
Here, 
\begin{eqnarray}
\la x\ra &=& \la x_0\ra~e^{-k\tau}+Ae^{-k\tau}\int_0^\tau dte^{kt}\sin(\omega t+\phi) \nn\\
&=& \frac{A\sin\phi}{k}~e^{-k\tau}+\frac{A}{k^2+\omega^2}[k\sin(\omega \tau+\phi) 
-\omega\cos(\omega \tau+\phi)\nn\\
&&\hspace{3cm}+e^{-k\tau}(\omega\cos\phi-k\sin\phi)].\nn\\
\end{eqnarray}
Using the above relations, $\Delta s_{tot}$ becomes 
\begin{eqnarray}
\la\Delta s_{tot}\ra &=& \frac{\la W\ra}{T}+\frac{A}{T}\left(\la x\ra\sin(\omega \tau+\phi)-\la x_0\ra\sin\phi\right)-\frac{k}{2T}(\la x\ra^2-\la x_0\ra^2)\nn\\
&=& \frac{\la W\ra}{T}-\frac{A^2\omega^2e^{-2k\tau}}{2k(k^2+\omega^2)^2T}(\omega\sin\phi+k\cos\phi)^2 \nn\\
&&+ \frac{A^2e^{-k\tau}\omega^2}{k(k^2+\omega^2)^2T}(k\cos\phi+\omega\sin\phi)[k\cos(\omega \tau+\phi)+\omega\sin(\omega \tau+\phi)]\nn\\
&&+\frac{A^2\cos 2\phi}{4kT}-\frac{A^2k(k^2+3\omega^2)}{4(k^2+\omega^2)^2T}\cos(2\omega \tau+2\phi)\nn\\
&&-\frac{A^2\omega^2}{4k(k^2+\omega^2)^2T}[k^2+\omega^2+2k\omega\sin(2\omega \tau+2\phi)].\nn\\
\label{exprstot}
\end{eqnarray}

Again, $\Delta s_{tot}$ for the given particular thermal distribution becomes linear in $x$ and hence it becomes Gaussian and is given by
\begin{equation}
P(\Delta s_{tot})=\frac{1}{\sqrt{2\pi\sigma_s^2}}\exp\left[-\frac{(\Delta s_{tot}-\la \Delta s_{tot}\ra)^2}{2\sigma_s^2}\right],
\end{equation}
where $\sigma_s^2=2\la\Delta s_{tot}\ra$. The Gaussian distribution $P(\Delta s_{tot})$ along with the above obtained equation for variance implies the validity of the detailed fluctuation theorem, which is only valid for the linear potential in the transient case, provided the initial distribution is canonical \cite{lah09}.

\subsection{Heat distributions}

Unlike $W$, $W_c$ and $\Delta s_{tot}$, the heat is not linear in $x$ (as shown below) and hence the distribution is not Gaussian. $Q$ is given by
\begin{eqnarray}
Q \equiv W-\Delta U &=& -\int_0^\tau dt ~x(t)\frac{df(t)}{dt}- \frac{1}{2}k(x^2-x_0^2)\nn\\
&& +xf(\tau)-x_0f(0).
\end{eqnarray}
The distributions for $Q$ cannot be obtained analytically. Hence we resort to calculating $Q$ using numerical simulation. However, it may be noted that the Fourier Transform of $P(Q)$ can be obtained analytically \cite{zon02,zon04}.

\section{Results and discussions}

\begin{figure}
\subfigure[]{\includegraphics[width=8cm]{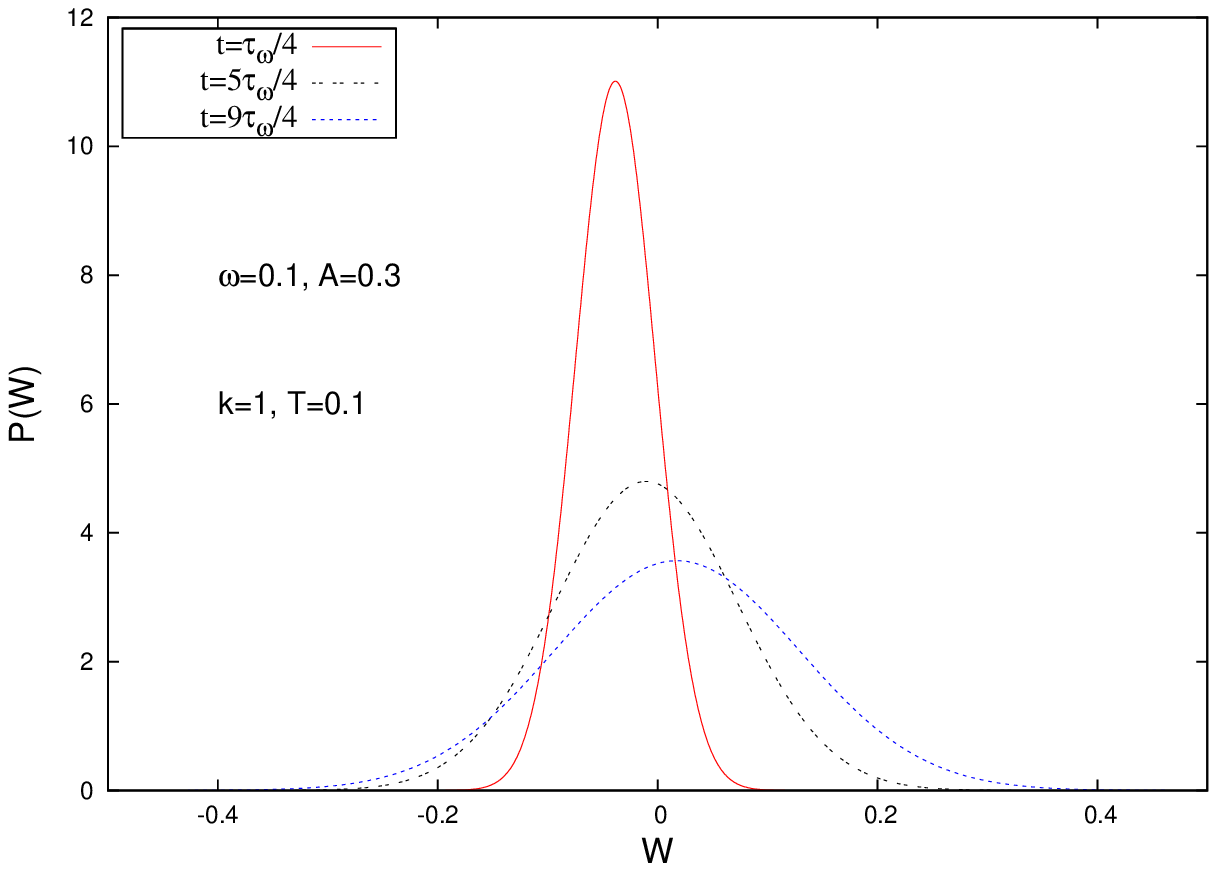}\label{anaW}}
\hfill
\subfigure[]{\includegraphics[width=8cm]{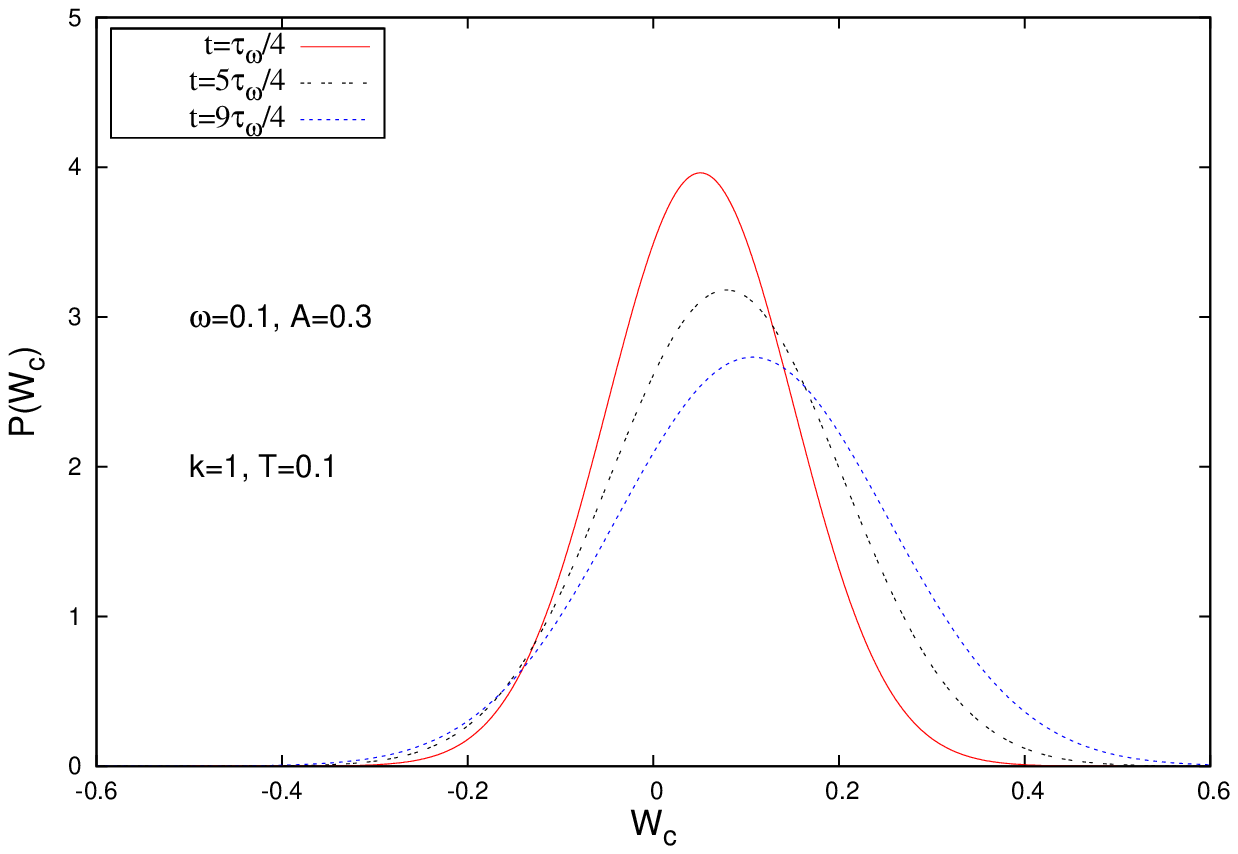}\label{anaWc}}
\subfigure[]{\includegraphics[width=8cm]{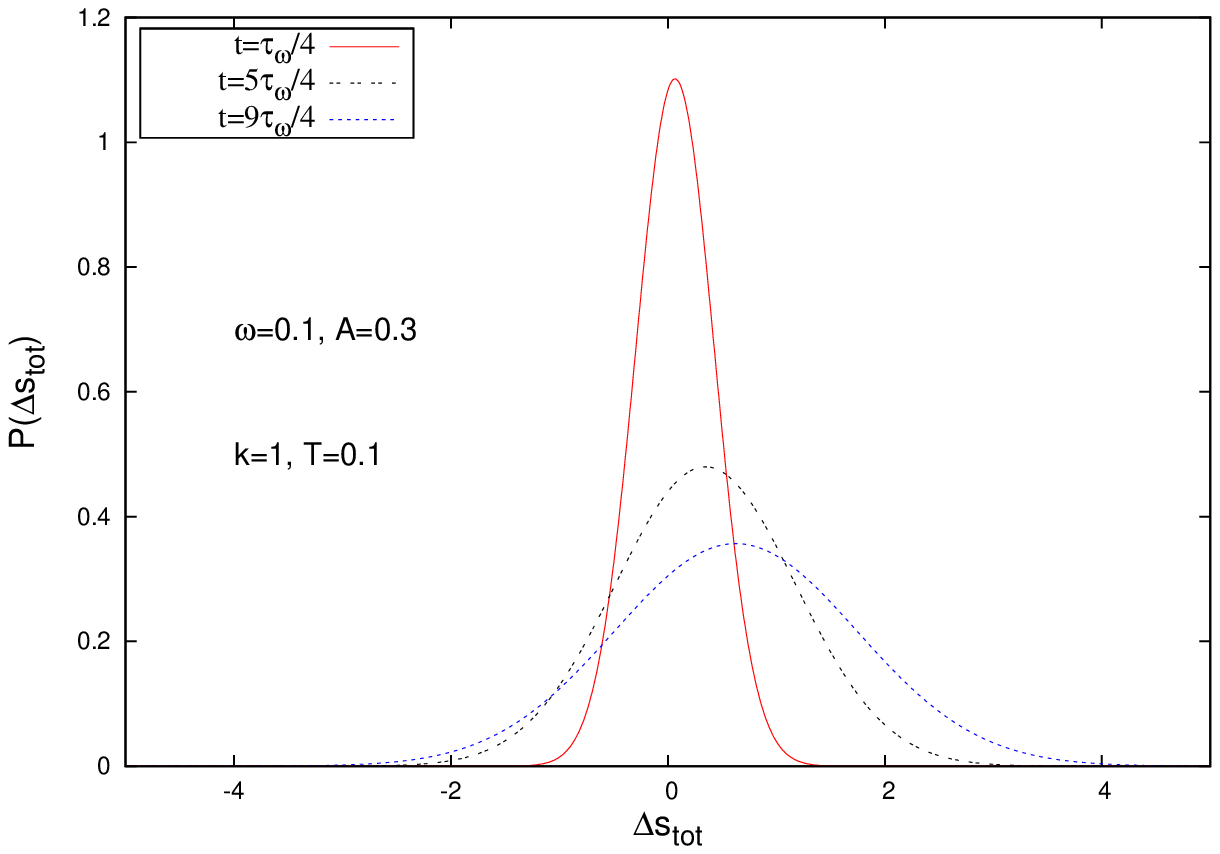}\label{anaSt}}
\hfill
\subfigure[]{\includegraphics[width=8cm]{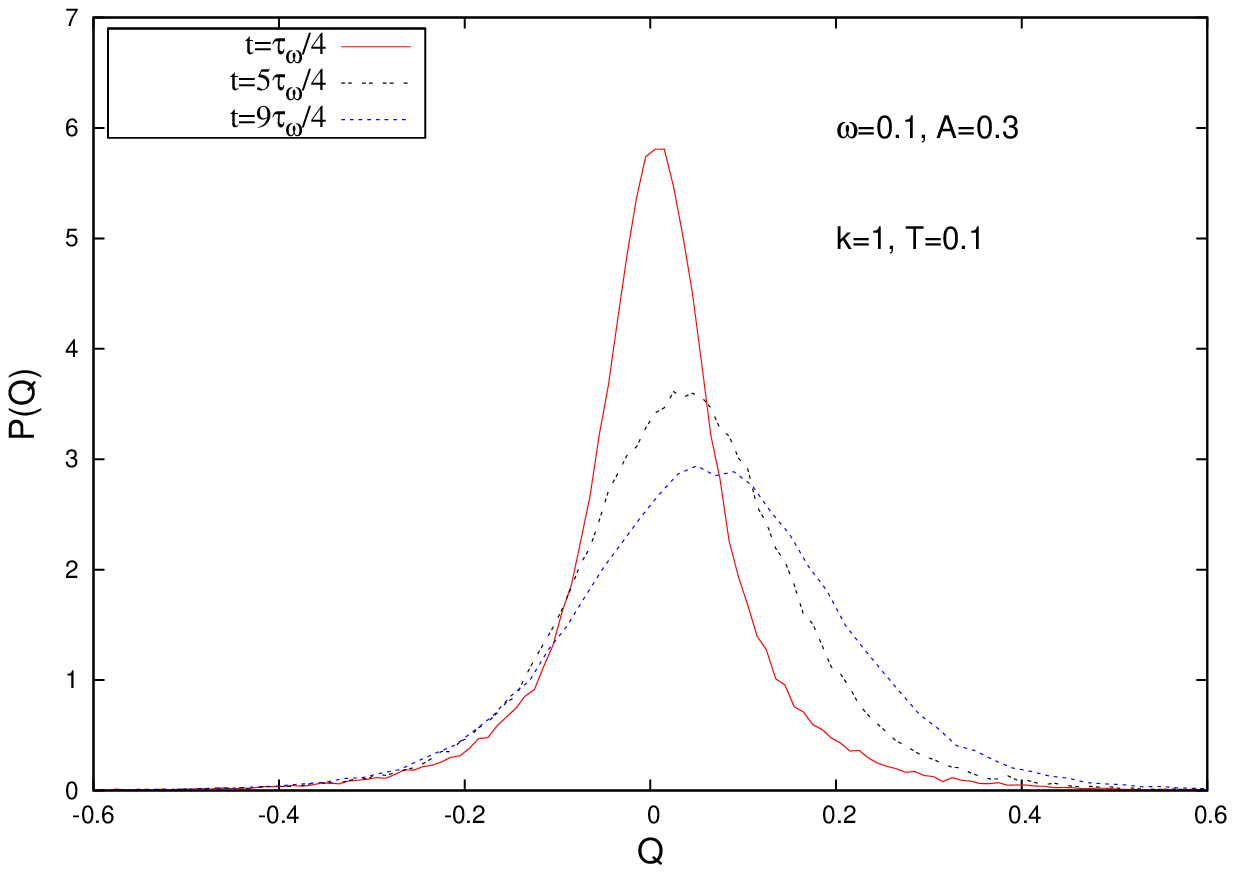}\label{simQ}}
\caption{(a) Analytical distributions of thermodynamic work of different times of observation. (b) Analytical distributions of classical work of different times of observation. (c) Analytical distributions of total entropy for different observation times. (d) Distribution of dissipated heat for harmonic potential for different values of observation time.} 
\end{figure}

In figures \ref{anaW}, (b), (c) and (d) we have plotted the probability densities for $W$, $W_c$, $\Delta s_{tot}$ and $Q$ respectively for different values of time $\tau=\tau_\omega/4$, $5\tau_\omega/4$ and $9\tau_\omega/4$, as indicated in the these figures. Here $\tau_\omega$ is the period of the external drive, namely, $\tau_\omega=2\pi/\omega$. We have taken all the physical quantities in dimensionless forms and have taken $\phi=0$ throughout our analysis. As anticipated, $W$, $W_c$ and $\Delta s_{tot}$ distributions are Gaussian and their most probable values (peaks in the distributions) are positive and shifts towards right with increase in time of observation. The distribution for $Q$ is non-Gaussian (figure \ref{simQ}), as anticipated from the previous discussion. In all these distributions, we find finite probability for the values of physical quantities that are atypical ($W<\Delta F$, $W_c<0$, $\Delta s_{tot}<0$ and $Q<0$). These trajectories are sometimes referred to as the transient second law violating trajectories. Fraction of such trajectories decrease as a function of the time of observation. The analytical expressions for such trajectories can be found out. For the thermodynamic work $W$, this fraction $f_W$ is obtained by integrating $P(W)$ from $-\infty$ to $\Delta F$. Similarly, $f_{W_c}$ and $f_{\Delta s_{tot}}$ are obtained by respectively integrating $P(W_c)$ and $P(\Delta s_{tot})$ from $-\infty$ to 0. The analytical results are 

\begin{numparts}

\begin{equation}
f_W = \frac{1}{2}\mbox{erfc}\left(\frac{1}{2}\sqrt{\frac{\la W\ra-\Delta F}{T}}\right);
\label{fW}
\end{equation}

\begin{equation}
f_{W_c} = \frac{1}{2}\mbox{erfc}\left(\frac{1}{2}\sqrt{\frac{\la W_c\ra}{T}}\right);
\label{fWc}
\end{equation}

\begin{equation}
f_{\Delta s_{tot}} = \frac{1}{2}\mbox{erfc}\left(\frac{1}{2}\sqrt{\la\Delta s_{tot}\ra}\right).
\label{fSt}
\end{equation}

\end{numparts}
These have the general form $f=\frac{a}{\sqrt{\tau}}e^{-c\tau}$ in the time-asymptotic limit. Further, it follows from the above equations that the fraction of violations cannot cross 0.5.

\begin{figure}
\centering
\includegraphics[width=8cm]{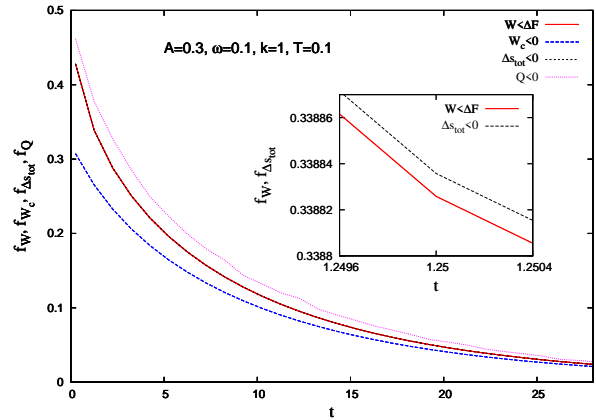}
\caption{Fraction of atypical trajectories for harmonic potential, as a function of time of observation (in units of $\tau_\omega$). The inset shows a magnified portion of the plots of $f_W$ and $f_{\Delta s_{tot}}$.}
\label{harmviols}
\end{figure}

\begin{figure}
\centering
\includegraphics[width=8cm]{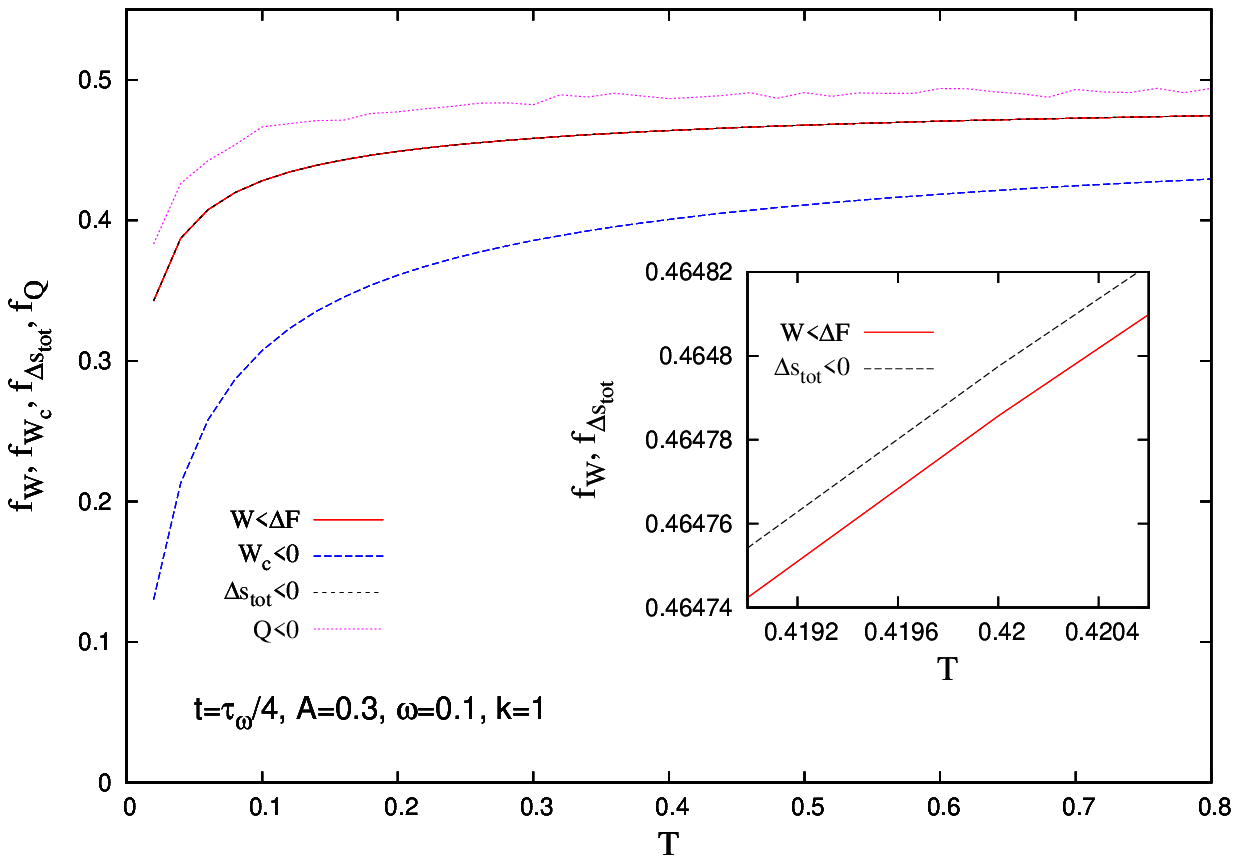}
\caption{Fraction of atypical trajectories for harmonic potential as a function of temperature. The inset shows a magnified portion of the plots of $f_W$ and $f_{\Delta s_{tot}}$.}
\label{harmviols_T}
\end{figure}

In figure \ref{harmviols}, we have plotted the fraction of  transient second law violating trajectories for $W$, $W_c$, $\Delta s_{tot}$ and $Q$. The first three are plotted using our analytical results. $f_Q$ for $Q$ has been obtained from numerical simulation. For this linear problem, we observe some systematics, i.e., we note that the number of atypical trajectories are larger for $Q$ and followed by the plots of $f_{\Delta s_{tot}}$ and $f_W$ which are seemingly coincident, and the lowest plot is that of $f_{W_c}$. However, the inset shows a magnified part of the plots of $f_{\Delta s_{tot}}$ and $f_W$, where the former is observed to be slightly higher than the latter. This extremely small difference is observed in the parameter regime $A, \omega\ll 1$. By analysis of the explicit expression (\ref{stot}), it can be shown that in this regime, $\Delta s_{tot}\approx (\la W\ra-\Delta F)/T$. Thus the violations become nearly same, which follows from equations (\ref{fW}) and (\ref{fSt}). The difference can be made large by changing the parameters. However, plotting all the four curves on the same graph does not bring clarity, but the conclusions remain unchanged. The systematics mentioned above for the fractions of atypical trajectories is a feature of the linear problem only. This can be noted from the next section. As the time of observation increases, the number of atypical trajectories decreases exponentially and goes to zero, thus leading to the well-defined classical thermodynamic behaviour. We would like to emphasize that given a trajectory, if it violates the relation $W\ge\Delta F$, this need not imply that the same trajectory will violate $W_c\ge 0$ or $\Delta s_{tot}\ge 0$ or $Q\ge 0$. This also follows from the simple fact that the fraction of violating trajectories is different for different physical quantities, and the trend seen in figure \ref{harmviols} remains same as a function of temperature, which has been plotted in figure \ref{harmviols_T}. Here too, the curves do not cross.  The above general observations hold for the case of a linear model. Some of these observations change qualitatively for non-linear systems, which we study below.

\begin{figure}
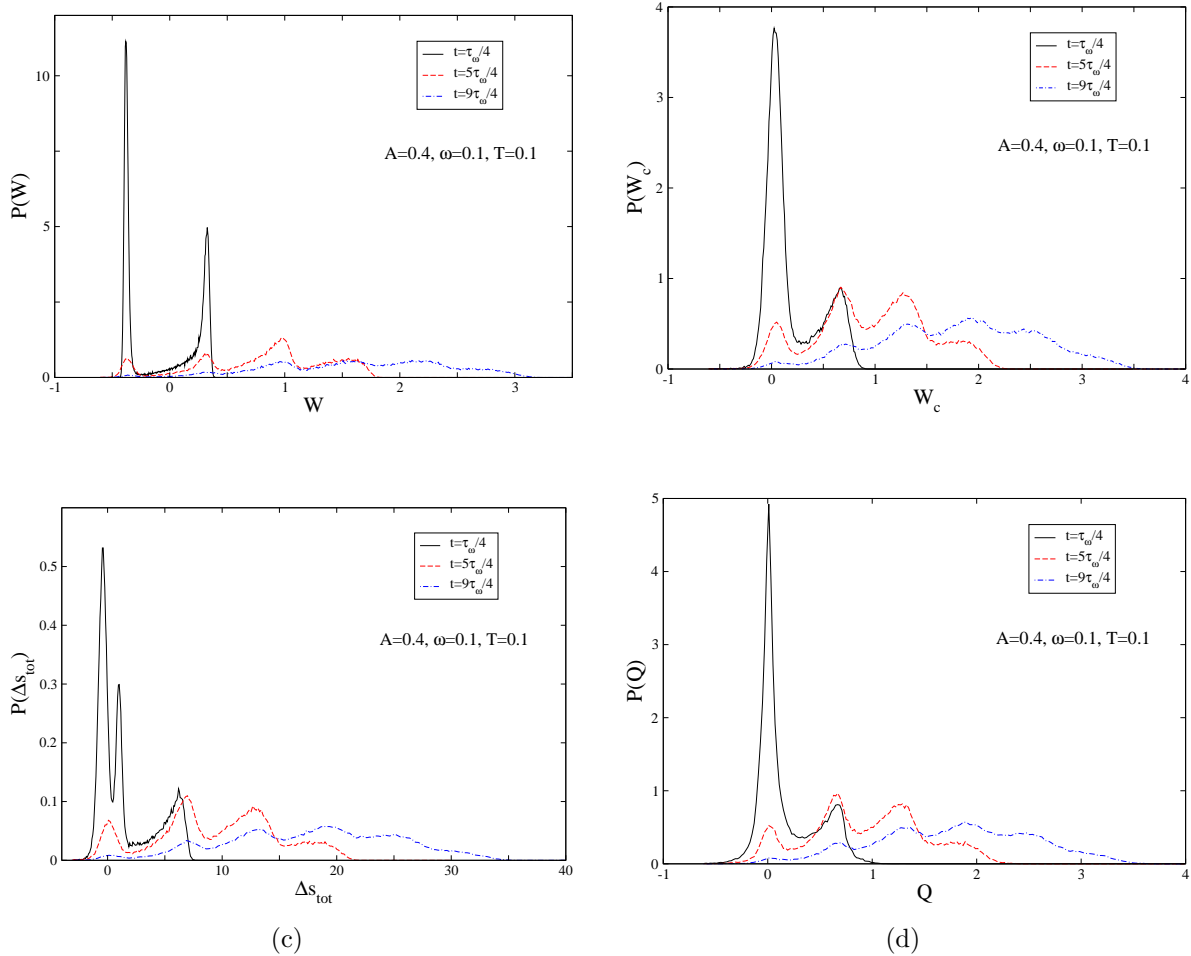

\subfigure[]{\includegraphics[width=7.5cm]{fig4a.eps}\label{MarW}}
\hfill
\subfigure[]{\includegraphics[width=7.5cm]{fig4b.eps}\label{MarWc}}
\subfigure[]{\includegraphics[width=7.5cm]{fig4c.eps}\label{MarQ}}
\hfill
\subfigure[]{\includegraphics[width=7.5cm]{fig4d.eps}\label{MarSt}}
\caption{(a) Distribution of thermodynamic work for double well for different values of observation time. (b) Distribution of classical work for double well for different values of observation time. (c) Distribution of total entropy for double well for different values of observation time. (d) Distribution of dissipated heat for double well for different values of observation time. }
\end{figure}

\begin{figure}
\centering
\includegraphics[width=8cm]{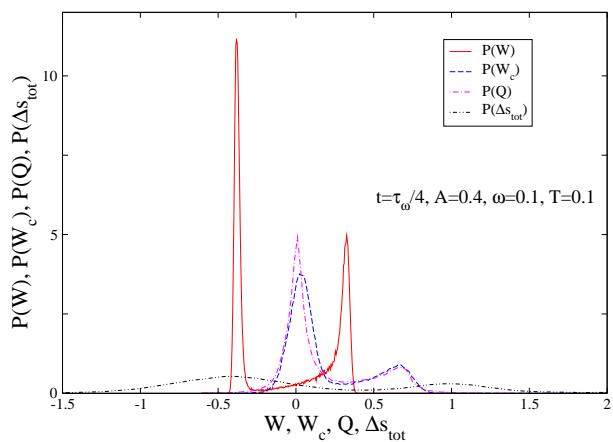}
\caption{Distributions for thermodynamic work, classical work, dissipated heat and total entropy for double well.}
\label{Mardists}
\end{figure}

\begin{figure}
\centering
\includegraphics[width=8cm]{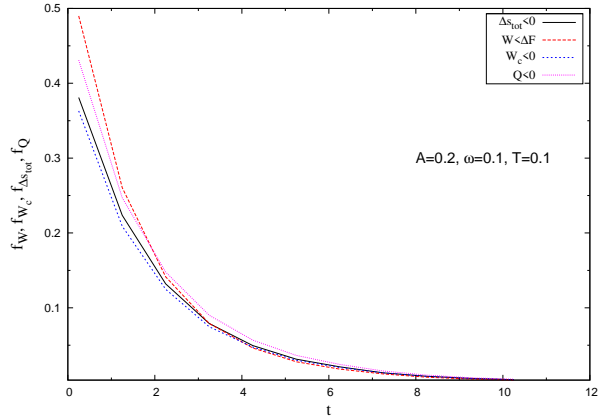}
\caption{Fraction of atypical trajectories in double well potential, as functions of time of observation (in units of $\tau_\omega$).}
\label{Marviols}
\end{figure}

\begin{figure}
\centering
\includegraphics[width=8cm]{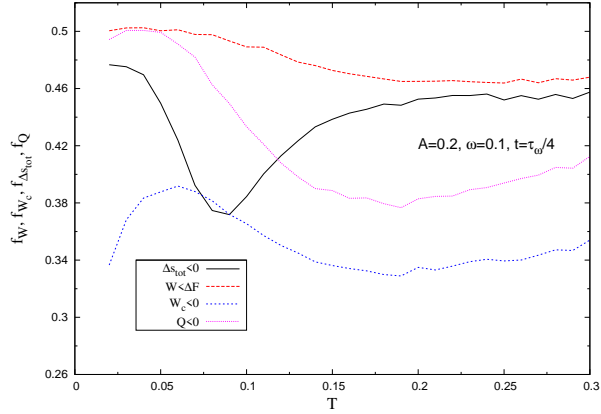}
\caption{Fraction of atypical trajectories in double well potential, as functions of temperature .}
\label{Marviols_T}
\end{figure}

\section{Non-linear model}

We now study the dynamics of a Brownian particle in a nonlinear system. The potential is of the form \cite{mar09}
\begin{equation}
V(x)=e^{-\alpha x^2}+\frac{k}{q}|x|^q.
\end{equation}
This particular potential has been studied in connection with stochastic resonance (SR) and is exhibits SR only for $q>2$ and for $q\le 2$, it does not exhibit SR. For our analysis, we have taken $q=4$ and $k=1$. All the physical quantities have been obtained numerically using the Heun's method. In figure \ref{MarW}, (b), (c) and (d) we have plotted distributions $P(W)$, $P(W_c)$, $P(\Delta s_{tot})$ and $P(Q)$. All the parameters are given in the figure. Initially the system is in equilibrium. In all figures 
we have multipeaked distributions due to intrawell and interwell dynamics. The protocols are applied for $t=\tau_\omega/4$, $5\tau_\omega/4$ and $9\tau_\omega/4$. The number of peaks in the distribution increase by two in each successive case. This corresponds to additional crossings of the particles over the intervening potential barrier. This not only broadens the distribution but its centre of mass or mean shifts to the right. The peak corresponding to the initial well motions \cite{sai07,mam08} shrink fast. The change in the free energy $\Delta F$ of the system is the difference between the free energies at the final and the initial values of the protocol, and is calculated by using the Jarzynski equality. We get $\Delta F=-0.31$ for $A=0.4$. We also obtain the values $\la e^{-W_c}\ra=1.00$ and $\la e^{-\Delta s_{tot}}\ra=1.07$, consistent with equations (\ref{IFT}) and (\ref{BKR}), respectively, within our numerical precision. Our protocol,as mentioned in the beginning, is a sinusoidal perturbation applied over different observation times as mentioned in the plots. It is clear that in all the cases, 
the fraction of atypical trajectories decreases as we increase the time of observation. Interestingly, $P(W)$ and $P(\Delta s_{tot})$ for $t=\tau/4$ exhibit the most probable value in the atypical region. This is quite interesting. 
However, it is to be noted that $\la W\ra$ and $\la\Delta s_{tot}\ra$ are greater than $\Delta F$ and zero, respectively. That is the typical behaviour at the macroscopic scale. In figure \ref{Mardists}, all the distributions are plotted on the same graph for comparison. The parameters are mentioned in the figure. We clearly see that $P(W)$ and $P(\Delta s_{tot})$ exhibit the peak corresponding to the most probable value in the region of atypical values. This is not a generic observation. This depends crucially on the values of the parameters, the time of observation and the protocol. 
However, we have verified with various protocols that the peak corresponding to the most probable value for $W_c$ and $Q$ always lie in the typical region. Average values of all the physical quantities increase with the time of observation as they are extensive in nature, as in the linear case. In figure \ref{Marviols}, we have plotted the fractional violations $f_W$, $f_{W_c}$, $f_{\Delta s_{tot}}$ and $f_Q$ as functions of the time of observation. These fractions decrease exponentially in time and coalesce time asymptotically, as in the linear case.
Again, unlike the linear model (wherein $f_Q>f_{\Delta s_{tot}}>f_{W}>f_{W_c}$), this trend is not maintained. Even two curves may cross, as shown in figure \ref{Marviols}. This is quite clear from the observations in figure \ref{Marviols_T}, where we have plotted fractional violations versus temperature. We can clearly see that  with increase in temperature, the plots cross each other. 

\section{Conclusions}

In our work, we have studied fluctuations in physical quantities such as thermodynamic work, classical work, total entropy and heat when the system is driven out of equilibrium. Our treatment is based on stochastic thermodynamics which gives prescription to calculate $W$, $W_c$, $\Delta s_{tot}$, $Q$, etc., for a given trajectory followed by the particle during evolution. Since we consider ensemble of trajectories, all the physical quantities become fluctuating variables, i.e., they take on random values depending on the trajectory of the particle. Unlike in thermodynamics, these physical quantities take on well-defined probability distributions which in turn satisfy the fluctuation theorems. We have analytically calculated $P(W)$, $P(W_c)$ and $P(\Delta s_{tot})$ for a sinusoidally driven linear system. The glaring differences between the linear and the nonlinear systems have been pointed out. Systematics which are present in linear systems are absent in nonlinear systems. These results are amenable to experimental verifications. We would also like to emphasize that in both the linear and in the nonlinear models, we observe that the fraction of trajectories violating typical behaviour are not greater than 50\%.

\ack{One of us (A.M.J) thanks DST, Government of India for financial support.}

\appendix

\section{Expressions for $\la W\ra$ and $\la W_c\ra$ for harmonic potential under sinusoidal driving}\label{appA}

\begin{eqnarray}
\la W\ra &=& \frac{A^2k}{4(k^2+\omega^2)}[\cos(2\omega \tau+2\phi)-\cos(2\phi)] + \frac{A^2\omega^2\tau}{2(k^2+\omega^2)} \nn\\ 
&&+ \frac{A^2\omega}{4(k^2+\omega^2)}[\sin(2\omega \tau+2\phi)-\sin(2\phi)]\nn\\ 
&&+ \frac{A^2\omega e^{-k\tau}}{(k^2+\omega^2)^2}[\omega\sin(\omega \tau+\phi)-k\cos(\omega \tau+\phi)](k\sin\phi-\omega\cos\phi)\nn\\ 
&&+ \frac{A^2\omega}{(k^2+\omega^2)^2}(k\sin\phi-\omega\cos\phi)(k\cos\phi-\omega\sin\phi)\nn\\ 
&&- \frac{A^2\omega\sin\phi}{k^2+\omega^2}[e^{-k\tau}\{\omega\sin(\omega \tau+\phi)-k\cos(\omega \tau+\phi)\}+k\cos\phi-\omega\sin\phi].\nn\\
\label{exprW}
\end{eqnarray}

\begin{eqnarray}
\la W_c\ra &=& \frac{A^2k}{4(k^2+\omega^2)}[\cos(2\omega \tau+2\phi)-\cos(2\phi)] + \frac{A^2\omega^2\tau}{2(k^2+\omega^2)} \nn\\
&&+ \frac{A^2\omega}{4(k^2+\omega^2)}[\sin(2\omega \tau+2\phi)-\sin(2\phi)]\nn\\ 
&&+ \frac{A^2\omega e^{-k\tau}}{(k^2+\omega^2)^2}[\omega\sin(\omega \tau+\phi)-k\cos(\omega \tau+\phi)](k\sin\phi-\omega\cos\phi)\nn\\ 
&&+ \frac{A^2\omega}{(k^2+\omega^2)^2}(k\sin\phi-\omega\cos\phi)(k\cos\phi-\omega\sin\phi)\nn\\ 
&&+ \frac{A^2k}{k^2+\omega^2}\sin^2(\omega \tau+\phi) - \frac{A^2\omega}{2(k^2+\omega^2)}\sin(2\omega \tau+2\phi)\nn\\ 
&&+ \frac{A^2e^{-k\tau}}{k^2+\omega^2}\sin(\omega \tau+\phi)[\omega\cos\phi-k\sin\phi]. 
\end{eqnarray}

\end{document}